\documentclass[review]{elsarticle}

\usepackage{geometry}
\usepackage[colorlinks,linkcolor=blue]{hyperref}
\usepackage{float}
\usepackage{graphicx}
\usepackage{subfigure}
\usepackage[utf8]{inputenc} 
\usepackage[T1]{fontenc}    
\usepackage{url}            
\usepackage{booktabs}       
\usepackage{amsfonts}       
\usepackage{nicefrac}       
\usepackage{microtype}      
\usepackage{lipsum}		
\usepackage{doi}
\usepackage{amsmath}
\usepackage{amssymb}
\usepackage{enumerate}
\usepackage[ruled,vlined]{algorithm2e}

\biboptions{authoryear}

\begin{document}

\begin{frontmatter}

\title{Urban Vehicle Mobility Characteristic Mining and Trip Generation Based on Knowledge Graph}

\author[mymainaddress]{Guilong Li}
\ead{liglong3@mail2.sysu.edu.cn}

\author[mymainaddress]{Yixian Chen}
\ead{chenyx96@mail2.sysu.edu.cn}

\author[mymainaddress]{Jun Xie}
\ead{xiej29@mail2.sysu.edu.cn}

\author[mymainaddress]{Qinghai Lin}
\ead{linqh8@mail2.sysu.edu.cn}

\author[mymainaddress]{Zhaocheng He\corref{mycorrespondingauthor}}
\cortext[mycorrespondingauthor]{Corresponding author}
\ead{hezhch@mail.sysu.edu.cn}

\address[mymainaddress]{Guangdong Provincial Key Laboratory of Intelligent Transportation System, School of Intelligent Systems Engineering,
Sun Yat-sen University, Guangzhou 510275, China}

\begin{abstract}

The operation of urban transportation produces massive traffic data, which contains abundant information and is of great significance for the study of intelligent transportation systems. In particular, with the improvement of perception technology, it has become possible to obtain trip data in individual-level of vehicles. It has finer granularity and greater research potential, but at the same time requires higher requirements in terms of data organization and analysis. More importantly it cannot be made public due to privacy issues. 

To handle individual-level urban vehicle trip big data better, we introduce the knowledge graph for the study. For organization of individual level trip data, we designed and constructed an individual-level trip knowledge graph which greatly improves the efficiency of obtaining data. Then we used the trip knowledge graph as the data engine and designed logical rules to mine the trip characteristics of vehicles by combining the transportation domain knowledge. Finally, we further propose an individual-level trip synthesis method based on knowledge graph generation to address the privacy issue of individual-level traffic data. The experiment shows that the final generated trip data are similar to the historical one in mobility patterns and vehicle associations, and have high spatial continuity.
\end{abstract}


\end{frontmatter}


\section{Introduction} \label{sec:introduction}

In recent years, the influx of a large number of people and the increase of private car ownership have intensified the pressure of urban transportation. On the other hand, urban transportation activities generate a large amount of data every day, making urban transportation a typical big data scenario. Especially with the improvement of collection devices and sensing technologies, data sensing equipment  such as automatic vehicle identification (AVI) detectors have been deployed broadly, making it possible to obtain individual-level trip data of vehicle. The individual-level trip data is naturally high-dimensional compared with the commonly used aggregate traffic data, and it contains richer information and has more potential power for research from the perspective of big data. Therefore, individual-level trip data is of great benefit to urban transportation management. However, the research of individual-level trip data faces some challenges in organization and analysis like the problem of scalability, fine granularity and data correlation. More importantly, the large-scale individual-level trip data cannot be made public due to data privacy issues, which greatly hinders related research. 

The inability to share individual-level urban trip data is a serious problem, and it can hinder the development of related research. Considering that the reason why individual-level trip data cannot be shared is that it involves the privacy of individuals, then the sharing problem can be solved if the privacy problem was solved. Based on this idea, we consider a de-privatization approach for individual-level trip data. Encrypting the individual's identity information is one approach that can be thought of, but if the trips are not processed, then they can still be mapped to individuals, so essentially trip privacy is not protected. Therefore deprivation of individual-level trip data needs to start from individual trips, i.e., to generate a new deprived data. Given that utility and privacy are competing factors, the de-privatized data is meaningless if it is not useful in most tasks. Therefore, the utility of the data is also considered in the process of acquiring the de-privatized data, such as the spatial and temporal patterns of trip. Unlike the generation of aggregated  data, individual-level trip data will involve trip associations between multiple individuals, which must be considered during the generation process, otherwise it may produce data that does not match the reality. Ensuring the correlation of trips between individuals is also a major difficulty in this task, so it is crucial to model the association information of trips between different individuals. Knowledge graphs are powerful for modeling realistic scenarios, especially the association between data, and it can preserve the information of individual trips to a great extent. Therefore, we consider using knowledge graphs to model individual-level trip data and investigate a "graph-generation" method to ensure the usability of the generated data by guaranteeing the similarity of the graphs and indirectly the similarity of the generated data with historical data.

In this paper, we introduced knowledge graphs to address the organization and privacy problem of individual-level trip data. Firstly, an individual-level trip knowledge graph was designed and constructed to organize the individual-level trip data, representing trip information as triples, which express the relation between data directly and significantly improves data acquisition efficiency. Based on the knowledge graph, a mobility characteristic mining method is proposed to analyze each vehicle's trip pattern. Each vehicle will get a ``label" according to trip characteristic by this method. Then all vehicles were classified into five groups. Finally, faced with the difficulty of obtaining individual-level trip data and the inability to open it up for privacy concerns, we proposed a trip knowledge graph generation algorithm. This algorithm can generate a trip knowledge graph where sensitive information is removed. The generated trip data has similarities with historical data in macro trip patterns, etc., and the dataset generated by this algorithm can be used publicly.

\section{Literature review} \label{sec:LR}

The concept of ``knowledge graph” with the modern meaning was proposed by Google in 2012, followed by further announcements of the development of knowledge graphs by institute of research or companies like Facebook\cite{noy2019industry}. In recent years, due to its powerful representation of data, domain knowledge graphs have also been gradually studied in various fields. Researchers have also started to model spatio-temporal traffic data using knowledge graph techniques.Zhang \cite{zhang2018structured}, and others constructed urban knowledge graphs for traffic accident inference using spatio-temporal correlation of traffic accident data with weather, road conditions and road networks. By dividing urban space into geographic entities, Yildirimoglu\cite{yildirimoglu2018identification} used trip data generated by different modes of transportation to correlate geographic entities  to construct a multi-layer traffic knowledge graph. 

The four-step planning scheme is a classical method for trip generation but scheme cannot meet the demand of advanced traffic management because it lacks the consideration of the time dimension and can only perform aggregate generation. The activity-based generation method is another primary trip generation method. This method considers that trip demand originates from activity demand and analyzes the micro-mechanisms of trip demand. Joubert et al. \cite{joubert2020activity} performed trip generation based on activity combined with the Bayesian approach and analyzed the reasons for trip generation. In the context of big data, the data-driven method of trip generation has arisen. This method focuses on data, discovers the association and characteristics of the data through the algorithm or model. Vaibhav Kulkarni et al. \cite{kulkarni2018generative} combined a neural network model (RNN) for trajectory generation.Kun Ouyang\cite{ouyang2018non} proposed a parameter-free human activity trajectory generation model.  Sánchez \cite{sanchez2011dealing} proposed a traffic flow prediction method based on the Bayesian networks. A Bwambale et al. \cite{bwambale2019modelling} proposed a method for generating set-meter-based trip demand based on cell phone data. Yang \cite{yang2020data} et al. used cell phone data and a location-based social network (LSBN) to achieve trip demand generation for residents using a random regression tree model.

For Knowledge graph generation, the mainstream method is the combination of machine learning method and deep learning model. For example, Simonovsky \cite{simonovsky2018graphvae} proposed a method towards small graph using variational autoencoders, while Cao  \cite{de2018molgan} and You \cite{you2018graphrnn} combine GAN and RNN respectively. Besides, You\cite{you2018graph} also proposed a graph convolutional policy for molecular graph generation. These methods have a common feature that the graphs are first parameterized, then learned by models, and finally the knowledge graphs are generated by decoding the parameters. These method is not suitable for transportation knowledge graphs for transportation field has special semantics, such as spatial continuity of trips, and they are difficult to learn well after parameterization.

\section{Methodology}
\subsection{Organization of  Individual-level Trip Data}\label{sec:organize data}
The commonly used structured data organization method has the following problems when dealing with individual-level trip data. 
Firstly, the acquisition efficiency of individual trip data in structured organization is $O(n)$, that is, with the increase of data scale, the efficiency of data acquisition decreases linearly. On the other hand, the rich correlation information between data can not be well expressed. To address this problem, we introduce knowledge graph to organize individual-level trip data, which has the following advantages.

\begin{itemize}
    \item The entities of the knowledge graph are unique and can be obtained by hashing. Therefore, if the structure is properly designed, the time complexity of individual-level trip data under the knowledge graph structure is $O(1)$.
    \item Knowledge graphs are able to express the long association of data, therefore the spatio-temporal correlation among the vehicles can be well indicated.
\end{itemize}

\begin{figure}[ht]
    \centering
    \includegraphics[width=9cm]{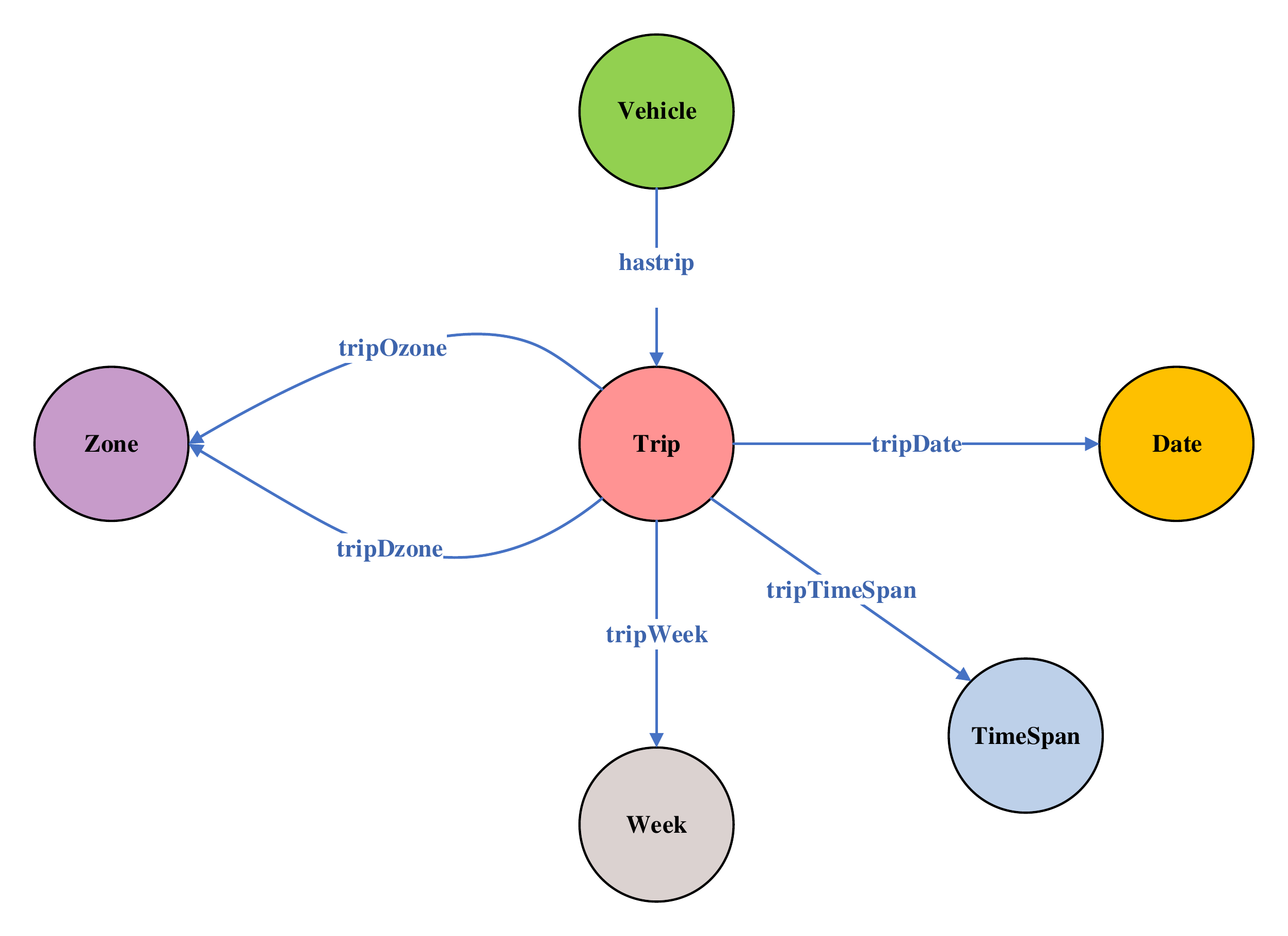}
    \caption{Structure of vehicle individual-level trip knowledge graph}
    \label{fig:structure_KG}
\end{figure}

Knowledge graph use entities and relations to describe data. For individual-level trip data, we designed the graph structure as shown in Figure \ref{fig:structure_KG}.

In Figure \ref{fig:structure_KG}, the circles represent entities, and the specific meanings represented by each type of entity are shown in Table \ref{tab:KG_entity}. Lines represent the different relations between entities with directions. The types of triples formed by various entities and relations and facts described by them are shown in Table \ref{tab:KG_triple}.

\begin{table}[h]
\centering
\caption{Entities of Vehicle Individual-Level Trip Knowledge Graph}
\label{tab:KG_entity}
\begin{tabular}{l|c}
\hline
\textbf{Entity}      & \multicolumn{1}{c}{\textbf{Meaning}}      \\ 
\hline

Vehicle & Unique identification of individual \\
Trip    & Abstract entities, representing one trip   \\
Week    & Nature of day, including working day and holiday   \\
TimeSpan & Time span of one day, e.g. Morning peak  \\
Date    & The specific time of trip   \\
Zone    & Traffic zone           \\ \hline

\end{tabular}
\end{table}

\begin{table}[h]
\centering
\caption{Triples of Vehicle Individual-Level Trip Knowledge Graph}
\label{tab:KG_triple}
\begin{tabular}{l|c}
\hline
\textbf{Triple}      & \multicolumn{1}{c}{\textbf{Fact}}      \\ 
\hline

(Vehicle)-[hastrip]-(Trip) & One trip was taken by the vehicle \\
(Trip)-[tripWeek]-(Week)    & Does the trip take place in workday or holiday   \\
(Trip)-[tripTimeSpan]-(TimeSpan)  & The time span of the trip  \\
(Trip)-[tripDate]-(Date) & The specific time of the trip  \\
(Trip)-[tripOzone]-(Zone)    & The origin traffic zone of the trip  \\
(Trip)-[tripDzone]-(Zone)    & The destination traffic zone of the trip           \\ \hline

\end{tabular}
\end{table}

The motivation of the above structure can be briefly described as abstracting each trip individual as an entity and abstracting each trip as an entity directly associated with the corresponding vehicle entity. The spatio-temporal information of the trip is associating with the trip entity. 

\subsection{Vehicle mobility characteristics mining}\label{sec: trip character}

The data are well organized and expressed in the individual-level trip knowledge graph, which contains lots of knowledge that can be mined, and mobility characteristics of vehicle is one of them. This section will propose a vehicle mobility characteristics mining method based on efficient data acquisition powered by the individual-level trip knowledge graph, through which three types of mobility characteristics can be obtained for each vehicle. On this basis, all vehicles are classified into five groups according to ``mobility label"  given by labeling system.

\subsubsection{Trip frequency characteristics mining}

 The characteristics of trip frequency reflect the frequency of the vehicle's trip. The method of extraction is to calculate the frequency of trips within a specified time range, and here we give four levels of trip frequency characteristics according to the frequency; see Table \ref{tab:frequency mining}.

\begin{table}[h]
\centering
\caption{The way to divide the frequency characteristics}
\label{tab:frequency mining}
\begin{tabular}{c|c}
\hline
Average daily trip frequency & Frequency characteristics   \\ \hline
frequency $\leqslant 0.16$      & Extremely low frequency  \\
$1.5\geqslant$ frequency $>0.16$ & Low frequency \\
$6.7>$ frequency $>1.5$ & General frequency \\
frequency $\geqslant 6.7$   & High frequency  \\ \hline

\end{tabular}
\end{table}

It is important to emphasize that trip frequency classification levels and thresholds vary according to different cities and demands, which should be adjusted according to data and requirements when extracting.

\subsubsection{Trip distribution characteristics mining}

The trip distribution characteristic reflects the extent of spatio-temporal aggregation of mobility. For example, some vehicles only travel at certain times of the day, and most of their trips are concentrated in one or several time spans, while others may spread over multiple time spans. Similarly, such difference also exists in spatial.  Trip distribution characteristics can be divided into the following items.
\begin{enumerate}
   \item[-] Trip temporal distribution
   \begin{enumerate}
        \item[$\bullet$] Trip time distribution
    \end{enumerate}
   \item[-] Trip spatial distribution
    \begin{enumerate}
        \item[$\bullet$] Trip origin distribution
        \item[$\bullet$] Trip destination distribution
    \end{enumerate}
\end{enumerate}

\textbf{``Highly concentrated"}, \textbf{``Concentrated"} and \textbf{``Dispersed"} are used for evaluate the concentration degree of each item, and the rules are as follows.
\begin{itemize}
    \item[(1)]  In equation \ref{eq:high_agg} $\left\lfloor x \right\rfloor$ denotes rounding to the nearest integer for $x$, $N$ denotes all time spans or traffic zones taken by the vehicle. and $n_i$ represents the number of trips in the $i$-th time span or traffic zone. If it is satisfied, then the concentration level is \textbf{``Highly concentrated"}.
    \item[(2)] If equation \ref{eq:high_agg} is not satisfied and equation \ref{eq:agg} is satisfied, the concentration is denoted as \textbf{``Concentrated"}. Otherwise, it is \textbf{``Dispersed"}.
\end{itemize}

\begin{equation}
    \sum_{i=1}^{\left\lfloor0.2\ast N\right\rfloor}{n_i\geq0.8\ast\sum_{i=1}^{N}n_i}\label{eq:high_agg}
\end{equation}

\begin{equation}
    \sum_{i=1}^{\left\lfloor0.3\ast N\right\rfloor}{n_i\geq0.7\ast\sum_{i=1}^{N}n_i}\label{eq:agg}
\end{equation}

With the above process, the trip distribution characteristics of vehicles will be expressed as different concentrations of the three items.

\subsubsection{Trip association characteristics mining}

Trip distribution characteristics can only measure the concentration of trips from a single dimension of time or space, while association characteristics further consider the spatial and temporal correlation of trips. Some individuals have a high probability of arriving at a traffic zone at a specific time, which reflects the strong spatio-temporal correlation of trips. Similar to the distribution characteristics, the association characteristics can be subdivided into the following three items.

\begin{enumerate}
   \item[-] Spatio-temporal correlation
   \begin{enumerate}
        \item[$\bullet$] Correlation of trip origin and time
        \item[$\bullet$] Correlation of trip destination and time
    \end{enumerate}
   \item[-] Spatial correlation
    \begin{enumerate}
        \item[$\bullet$] Correlation of Trip origin and destination
    \end{enumerate}
\end{enumerate}

In order to quantitatively evaluate the association of individual vehicle trips, the ``association score" of trips is defined, and it is calculated by Equation \ref{eq:rel_score}-\ref{eq:rel_score1}.

\begin{equation} 
    \frac{\sum_{i=1}^{n}{\min(1,\left(1+\frac{relu\left(q_i-m\right)}{q_i}\right)\ast\max{\left(p_{i,j}\right)})}}{\sum_{i=1}^{n}\sum_{j}^{m}p_{i,j}}\ast100 \label{eq:rel_score}
\end{equation}

\begin{equation}
    q_i=\max(1,\rho\sum_{j=1}^{m}{p_{i,j})}\label{eq:rel_score1}
\end{equation}

In equation \ref{eq:rel_score}, $p_{i,j}$ represents the number of trip for the $i$-th time span or traffic zone(if calculate Spatial correlation) with the $j$-th traffic zone as a combination; $n$ is the number of all trip time spans or zones the vehicle taken; $m$ is the number of zones associated with a specific time span or zone; $\rho$ is the reward factor, whose recommended value is between $0.2$ and $0.3$.

With the equation of association score, each association feature item corresponds to a $score\in [0,100]$, where a higher score represents a stronger association.

\subsubsection{Vehicle's label system}

According to above subsections, we can get the expression of three kinds of mobility characteristics.
This section will propose a label system based on mobility characteristics, which create a label for each vehicle. For the purpose of concise expression, the following two concepts are defined first.

\begin{figure}[h]
\setlength{\abovecaptionskip}{0.cm}
\centering
\includegraphics[width=12cm]{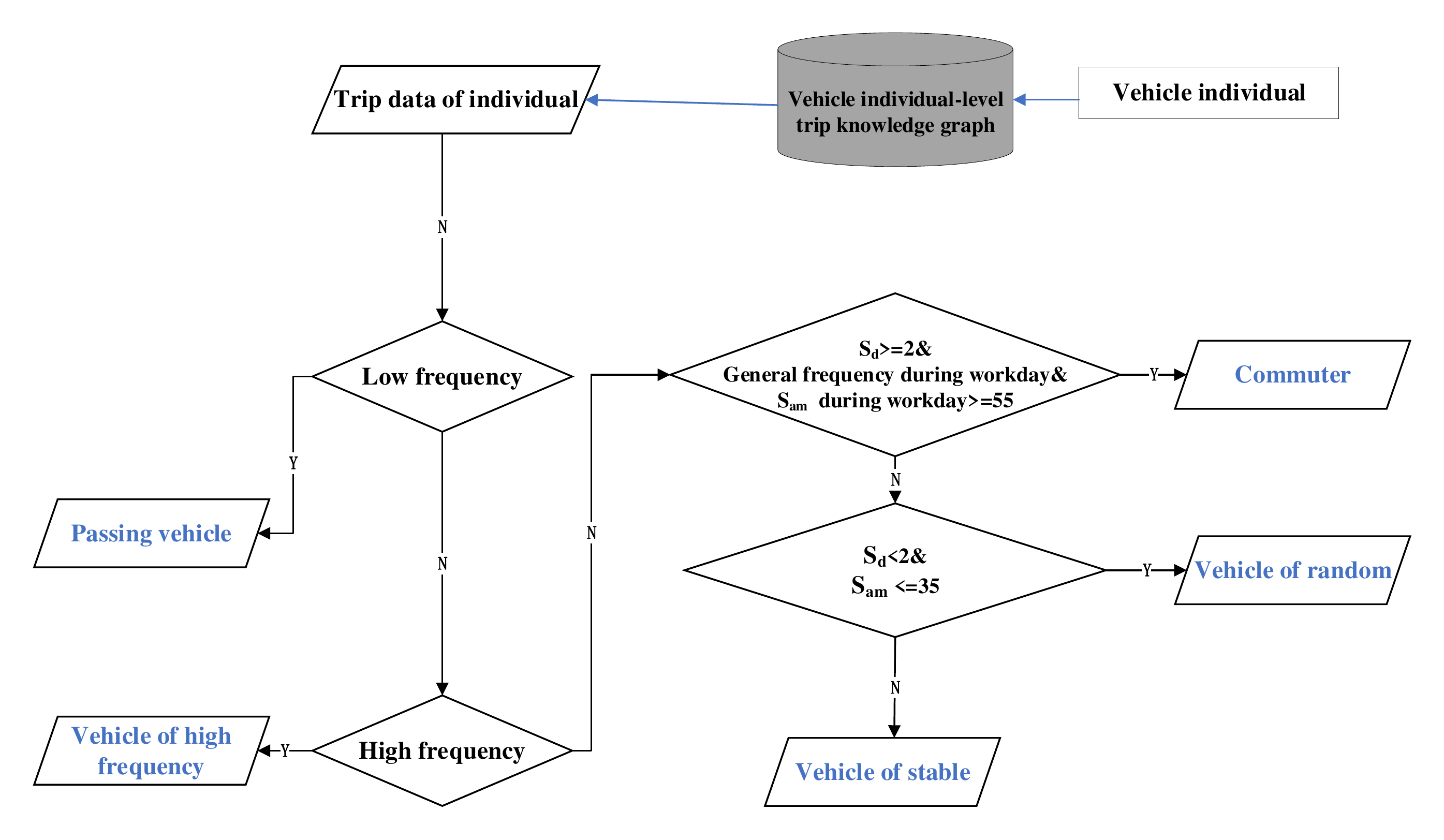}
\caption{Vehicle's labeling system.}
\label{fig:label system}
\end{figure}

\begin{itemize}
    \item[$\bullet$] \textbf{Distribution Concentration} $S_d$ denotes the concentration of ``highly concentrated", ``concentrated", and ``dispersed" as 2, 1, and 0 respectively. It is calculated as the sum of the concentrations corresponding to the three items of the distribution characteristics.
    \item[$\bullet$] \textbf{Associated Mean Score} $S_{am}$ is calculated as the mean of the scores of the three items of the association characteristic.
\end{itemize}
Based on the above definition, the flowchart of the vehicle's labeling is given in Figure \ref{fig:label system}. As shown, the vehicle is divided into five categories ``Passing vehicle", ``Commuter", ``Vehicle of stable ", ``Vehicle of random ", and ``Vehicle of high frequency " based on the scores of three types of mobility characteristics.

\subsubsection{Division of label subgraphs}\label{sec:division subgraph}

Through the mobility characteristics mining, each vehicle has a new label. It can be represented by the individual-level trip knowledge graph as knowledge by the following processing. 
\begin{itemize}
    \item[$\bullet$] Extract the mobility labels as a type of entities named $Label$ which contains five specific entities and add it to the individual-level trip knowledge graph.
    \item[$\bullet$] Add a new relation $TripType$ to associate the $Vehicle$ entity with $Label$ entity, and obtain a new triplet type ``$(Vehicle)-[TripType]-(Label)$".
\end{itemize} 

With the introduction of the $Label$ entities, the individual-level trip knowledge graph can be divided into five ``label subgraphs", in which the vehicle entities have the same $Label$ entity. The following generation problem on trip knowledge graph was based on each subgraph, which helps  retain the mobility characteristics of typical urban groups.  Similarly, we can extracte subgraph from individual-level trip knowledge graph that has the same $Date$ entity and we denoted it as ``date subgraph". 

\subsection{Trip generation based on knowledge graph}

The trip data are extremely valuable for ITS research, but they are also very high cost to collect. Therefore it is of great importance to investigate how to generate data for open and research purposes based on the available data. With the individual-level trip knowledge graph, we studied that how to perform knowledge graph generation on it, hence obtaining synthesis trip data by retrieving from the generated knowledge graph. 

The trip knowledge graph consists of the basic entities and the relations shown in Figure \ref{fig:structure_KG}. When the amount of data reaches a certain scale, the depth of association between entities is deeper and the ways of association are diversified, making the knowledge graph a ``complex graph" where entities are associated through multi-level and different types of relations. By exploring the triplet structure, we find that it can be decomposed into multiple ``unit graphs" that are associated by partial types of entities with defined association patterns.

\begin{figure}[h]
\setlength{\abovecaptionskip}{0.cm}
\centering
\includegraphics[width=12cm]{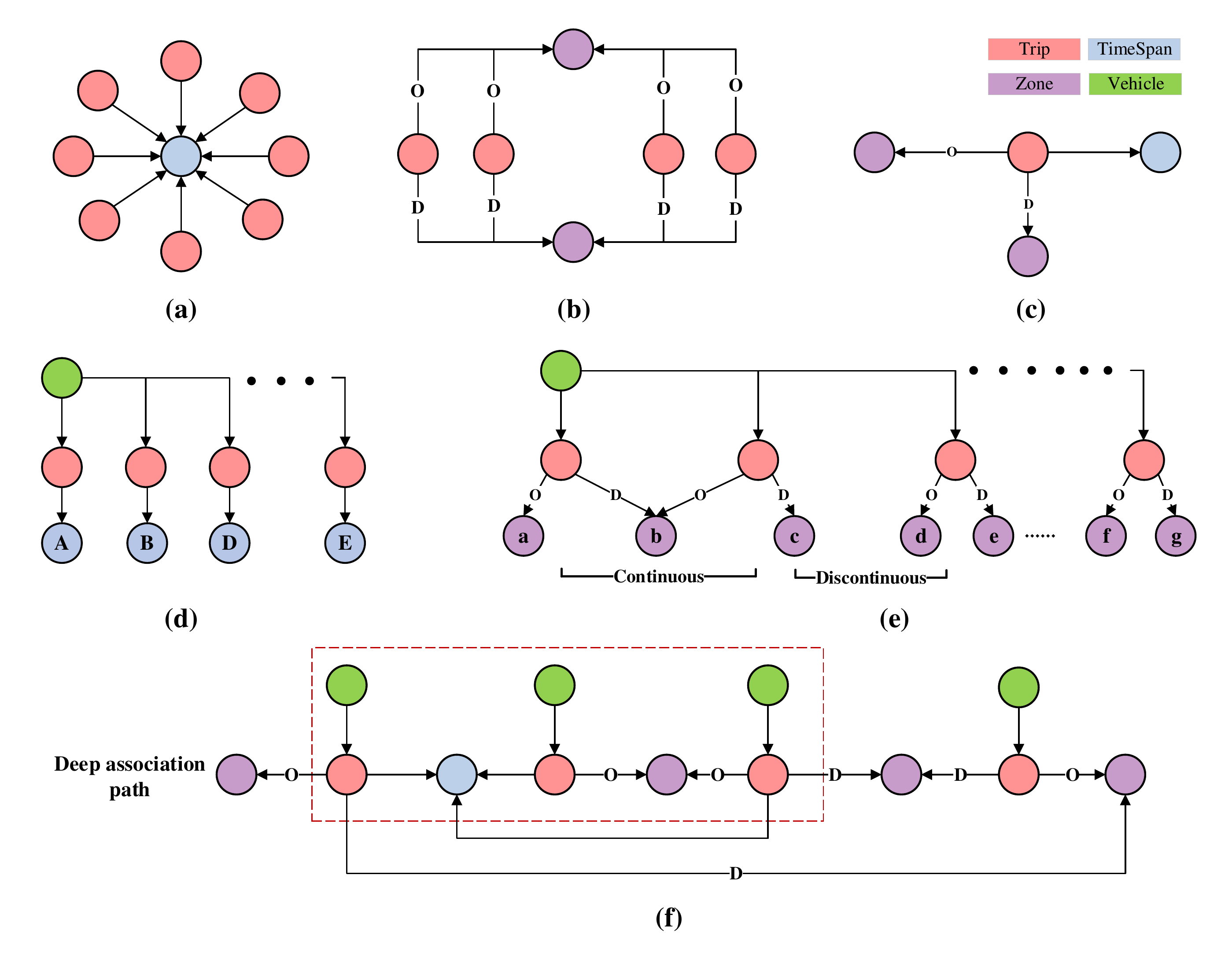}
\caption{The core unit graphs decomposed from the individual-level trip knowledge graph.}
\label{fig:unit structure}
\end{figure}
\subsubsection{Extraction of unit graphs from trip knowledge graph}

After analyzing all possible unit graphs and their structure meanings on mobility, we identified the core unit graphs as shown in Figure \ref{fig:unit structure} which can restore all information without information redundancy. It should be pointed out that since the $Week$ entity is a mapping of $Date$ entity, it is not considered when extracting unit graphs. Besides, the $Date$ is also not considered for all of them is same in date subgraph. Next we will analyze and profile these unit graphs in order of association depth. 

\paragraph{Unit graphs of one-level association}

These unit graphs are made up of one type of entity directly related to other category entities. We identified the core unit graphs among all of unit graphs of the first-level association which was shown in Figure \ref{fig:unit structure}(a)-(c). It is easy to see that they are all unit graphs containing the ``$Trip$" entity. (a) is the unit graph that only contains ``$(Trip)-[tripTimeSpan]-(TimeSpan)$" triple. This unit graph contains the information of vehicles' trip time, and the trip number of different time spans can be retrieve by it. So it reflects the temporal pattern of trips, as the unit graph of ``trip temporal pattern". Similarly, (b) is the unit graph containing only the trip origin and destination triples reflecting the trip spatial information, through which the trip number under different OD combinations can be obtained. Therefore, (b) is the unit graph of  "trip spatial pattern". For (c), it takes the trip entity as the core and considers both its temporal and spatial correlations, i.e., the unit graph describes the spatio-temporal information of a single trip. Each trip in the graph corresponds to a unit graph of this structure, and the set of such unit graphs is able to describe the overall spatio-temporal pattern of trips, with information covering both (a) and (b).
   
Through the above analysis, these core unit graphs of one-level association can be summarized as unit graphs that reflect \textbf{the spatio-temporal patterns of trips}.

\paragraph{Unit graphs of two-level association}

Figure \ref{fig:unit structure}(d)-(e) is the unit graphs of two-level association. It can be seen that vehicle individuals are included in them, so they reflect the trip information at the individual level. (d) contains the triples of ``$(Vehicle)-[hastrip]-(Trip)$" and ``$(Trip)-[tripTimeSpan]-(TimeSpan)$", which can reflect the trip number of the vehicle and the time spans taken by it for tripping. In other words, it reflects the combination and preference of  vehicles individual for the choice of trip time spans. For unit graph (e), it can be seen intuitively that it reflects the choice of the origin and destination of the vehicle, in addition, it also hides an important information, that is trip continuity.Assume that the $Trip$ entities in \ref{fig:unit structure}(e) are arranged from left to right according to the time of occurrence, then  we can judge whether the adjacent trips are continuous in space through the unit graph.As shown in (e), if the destination $D_f$ of the previous trip $T_f$ of the adjacent trip is the same as the origin $O_l$ of the next trip $T_l$, the space of the two trips is continuous, which shows that there is a path that ``$(Trip:T_f)-[tripDzone]-(Zone)-[tripOzone]-(Trip:T_l)$" in the unit graph. On the contrary, it shows that the two trips do not have spatial continuity. 

\paragraph{Unit graphs of the multi-level association}

Since the considered association levels can be different, it is impossible to find a standard structure for the multi-level association unit graph. Responding to this situation, we try to extract the key information reflected under multi-level association from the unit graph with shallow association depth, and to find a more concise way to describe.

Figure \ref{fig:unit structure}(f) is a unit graph of multi-level association. we can see that the four types of  entities are associated  through one-level or two-level or multi-level different kinds of relations. Through the analysis of this complex association, we found that the deep association path are all made by the $Trip$ entity through the correlation of $TimeSpan$ and $Zone$ entities. Through further analysis We found that this deep association is essentially a description of the association between vehicles through trips. In other words, we can restore this deep association relations by simply describing the association between vehicles. For example, we can describe the deep correlation in the red box by ``the vehicle in the middle trips with the other two vehicles at the same time span and at the same origin respectively". Therefore, the multi-level association of trip knowledge graph essentially reflects the \textbf{association information between vehicles}.

\subsubsection{Construction of characteristic graph based on unit graphs}
The analysis of the complete unit graphs shows that the information contained in the trip knowledge graph can be divided into four main categories, which are:
\begin{itemize}
    \item[$\bullet$] Trip spatio-temporal pattern.
    \item[$\bullet$] Trip temporal combination of vehicle.
    \item[$\bullet$] Trip spatial continuity of vehicle.
    \item[$\bullet$] Association between vehicle.
\end{itemize}
The spatial information of vehicle trips, i.e., the choice of trip origin and destination can be covered by the trip spatial continuity of vehicle which is more of the trip semantic describe. 

For the generated trip knowledge graph, we need to ensure its consistency with the original trip knowledge graph in terms of the above information and the reasonableness of its own. To solve this problem, we first carried out the extraction of characteristic graphs based on the corresponding unit graphs, which enables a more efficient representation of the information. Then we formally represent and model the characteristic graphs to obtain their mathematical models. In this section, the extraction and modeling of characteristic graphs are described in detail.

\begin{figure}[h]
\setlength{\abovecaptionskip}{0.cm}
\centering
\includegraphics[width=12cm]{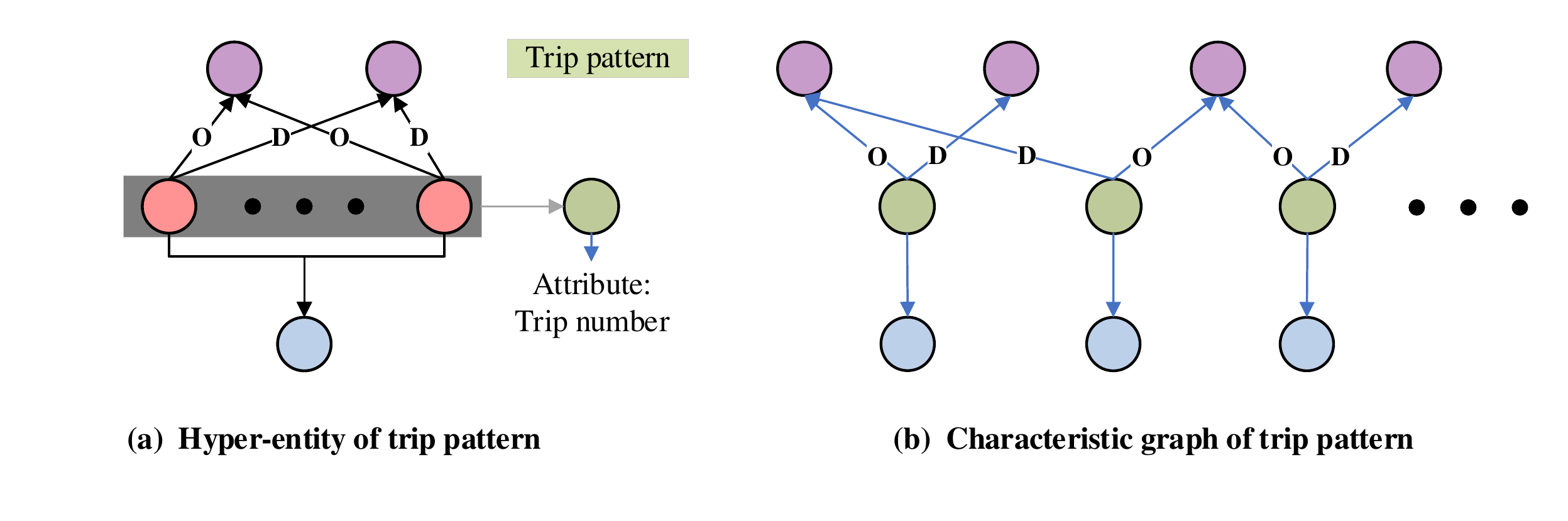}
\caption{Characteristic graph of trip spatio-temporal pattern.}
\label{fig:cha_graph_pattern}
\end{figure}

\paragraph{Characteristic graph of trip spatio-temporal pattern}
As described in \ref{sec:organize data}, in constructing the trip knowledge graph, each trip are abstracted as an $Trip$ entity. Denote the set of trip entities in the trip knowledge graph as $E_T$, then each $Trip$ entity in it can be extracted to get a spatio-temporal information unit graph as shown in Figure \ref{fig:unit structure}(c). If the corresponding $Trip$ entity of a specific trip and its time span occurred, origin and destination are represented by $T$, $S$,$Z_o$ and $Z_d$  respectively, then the general expression of the trip unit graph can be expressed as  ``$(Zone:Z_d)-[tripDzone]-(Trip:T)-[tripOzone]-(Zone:Z_o) \& (Trip:T)-[tripTimeSpan]-(TimeSpan:S)$". For each $Trip$ entity in the set $E_T$, extract its unit graph and record its  $S$,$Z_o$ and $Z_d$. On this basis, We aggregate the $Trip$ entities of $S$,$Z_o$,$Z_d$ which are all the same to obtain a type $Trip pattern$ of hyper-entity as shown in Figure \ref{fig:cha_graph_pattern}(a). In this process, the number of $Trip$ entities aggregated by each super-entity is recorded as its property. Then the characteristic graph of trip pattern can be constructed. This characteristic graph contains three types of entity as shown in \ref{fig:cha_graph_pattern}(b) of which the entity of $Trip pattern$ has a property of trip number aggregated and it retains the association of $Trip$ entities it aggregates with the $Zone$ and $TimeSpan$ entities. Therefore, in this characteristic graph, each hyper-entity represents a combination of ODT of vehicle trips in the city, i.e., a pattern of trips, and its frequency of occurrence is expressed by its property.

Assuming that there are a total of $N$ hyper-entities in the trip spatio-temporal pattern, we denote the $i$-th hyper-entity by $m_i$ and its properties by $r_i$. On this basis, this characteristic graph can be modeled with a discrete distribution $F$.

\begin{equation}
    F \sim p(X=m_i) = \frac{r_i}{\sum_{i=1}^N{r_i}}\label{eq:trip_pattern}
\end{equation}

\paragraph{Characteristic graph of trip temporal combination}

\begin{figure}[htb]
\setlength{\abovecaptionskip}{0.cm}
\centering
\includegraphics[width=12cm]{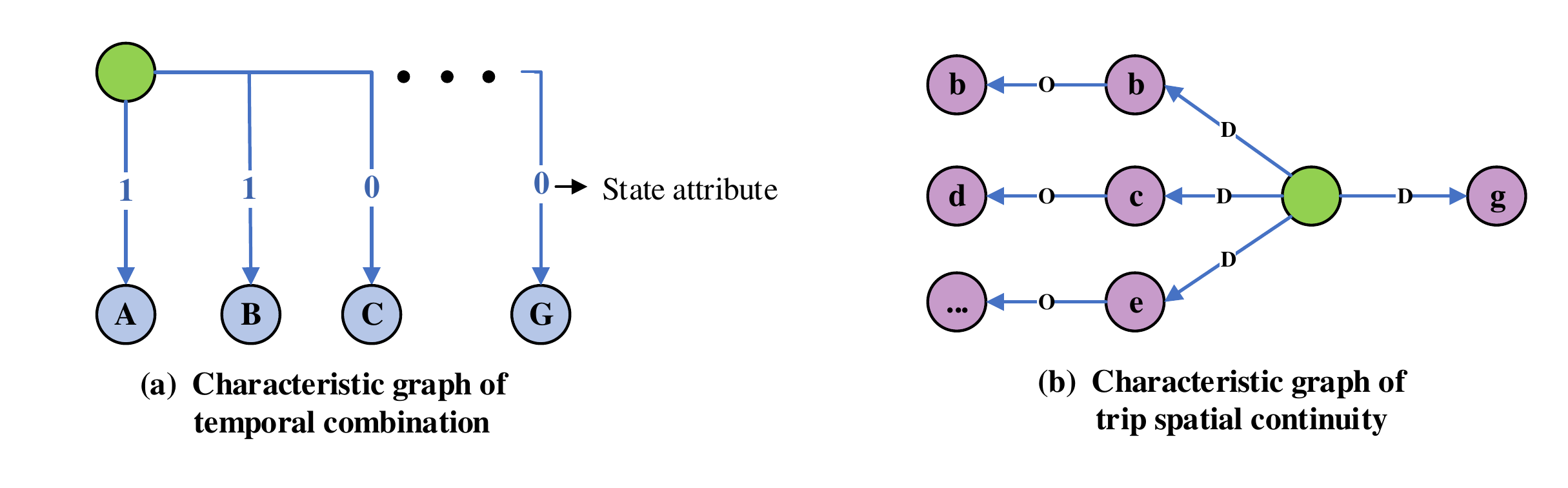}
\caption{Characteristic graph of temporal combination and trip spatial continuity.}
\label{fig:cha_graph_contiunity}
\end{figure}

For the unit graph shown in \ref{fig:unit structure}(d), it can express the number of trips of vehicles and the combination of the time spans the trips occurred. In the construction of the trip knowledge graph, only the trips that have occurred are expressed, while the time spans which the vehicle no trip are not expressed directly. In order to represent and model this information in a more standardized way, we constructed the characteristic graph of trip temporal combination as shown in Figure \ref{fig:cha_graph_contiunity}(a) based on the unit graph.

As shown in \ref{fig:cha_graph_contiunity}(a), it contains two types of entities, which are $Vehicle$ and $TimeSpan$. Contrary to the original trip knowledge graph, in this characteristic graph, the $Vehicle$ entity are associated with all entities of $TimeSpan$ through the relation of $ChooseTimeSpan$ type we defined. This relation has a status property to express whether a vehicle of the head entity has made a trip in the time span of the tail entity, $1$ if it has, and else $0$. In other words, if a path that $(Vehicle:V)-[hastrip]-(Trip:T)-[TripTimeSpan]-(TimeSpan:S)$ exists in the unit graph of vehicle, then the state attribute of $ChooseTimeSpan$ that associate $V$ and $S$ is $1$. 

For a characteristic graph of trip temporal combination of a specific vehicle $v_m$, we can represent it as a vector as shown in \ref{eq:T1}.

\begin{equation}\label{eq:T1}
    C(v_m) = [s_{m1},s_{m2},\cdots,s_{mL}]^\mathrm{T}
\end{equation}

Equation \ref{eq:T1} defines the mapping of vehicles to its vector expression of characteristic graph of trip temporal combination where the vector length $L$ corresponds to the number of entities of the $TimeSpan$ type. Each dimension of the vector in the equation  a time span. Denote the time span corresponds to $n$-th dimension in this vector as $t_n$. The value $s_mn$ in the vector is the value of the state attribute of the relation in the triple ``$(Vehicle:v_m)-[ChooseTimeSpan]-(TimeSpan:t_n)$" of the vehicle $v_m$ characteristic graph, which is obviously 0 or 1.

\paragraph{Characteristic graph of trip spatial continuity}

In the process of collecting trip data, there are many reasons for the discontinuity of vehicle's trip, such as missed detection. This information can be reflected by Figure \ref{fig:unit structure}(e). Different from other trip information, when performing the generation, we need to ensure the spatial continuity of vehicle's trips as much as possible rather than pursuing the similarity with the historical data for it is an objective fact.
In order to represent the spatial continuity of trip efficiently, the characteristic graph of trip spatial continuity is constructed as shown in Figure \ref{fig:cha_graph_contiunity}(b).  This characteristic graph expresses the spatial continuity of adjacent trips using the. It should be emphasized that the $Zone$ node of the graph is not unique, which is facilitate the calculation of continuity. Note the destination of the previous trip and the origin of next trip are $z_d^f$ and $z_o^l$ respectively. Then there are two type association path starting with $Vehicle$ entity. One is $(Vehicle)-[TripDzone]-(Zone:z_o^f)-[TripOzone]-(Zone:z_o^l)$, which we call ``long edge". The other is $(Vehicle)-[TripDzone]-(Zone:z_o^f)$, which we call ``short edge". 

From the characteristic graph of trip spatial continuity, the following two types of information can be obtained intuitively.
\begin{itemize}
    \item[$\bullet$] \textbf{Whether the two adjacent trips are spatial continuity.} Defining a indicator function $U$ to represent whether there is spatial continuity, then it can be judged by comparing whether two $Zone$ of ``long edge" are the same as shown in Equation \ref{eq:st_con1}. This facilitates the calculation of the vehicle trip spatial continuity.
    \item[$\bullet$] \textbf{The current position of the vehicle.} The $Zone$ entity in ``short edge" is the destination of the last trip of the vehicle, which is its current position. It was used in the graph generation logic.
\end{itemize}

\begin{equation}\label{eq:st_con1}
U(z_d^f,z_o^f)=
    \begin{cases}
    1 \qquad if \ z_d^f=z_o^f\\
    0 \qquad else
    \end{cases}
\end{equation}

\paragraph{Characteristic graph of vehicle association}

In the  trip knowledge graph, different kinds of associations can be generated between vehicles according to  the spatio-temporal information of trips, and these associations can be divided into the following two categories.

\begin{figure}[h]
\setlength{\abovecaptionskip}{0.cm}
\centering
\includegraphics[width=12cm]{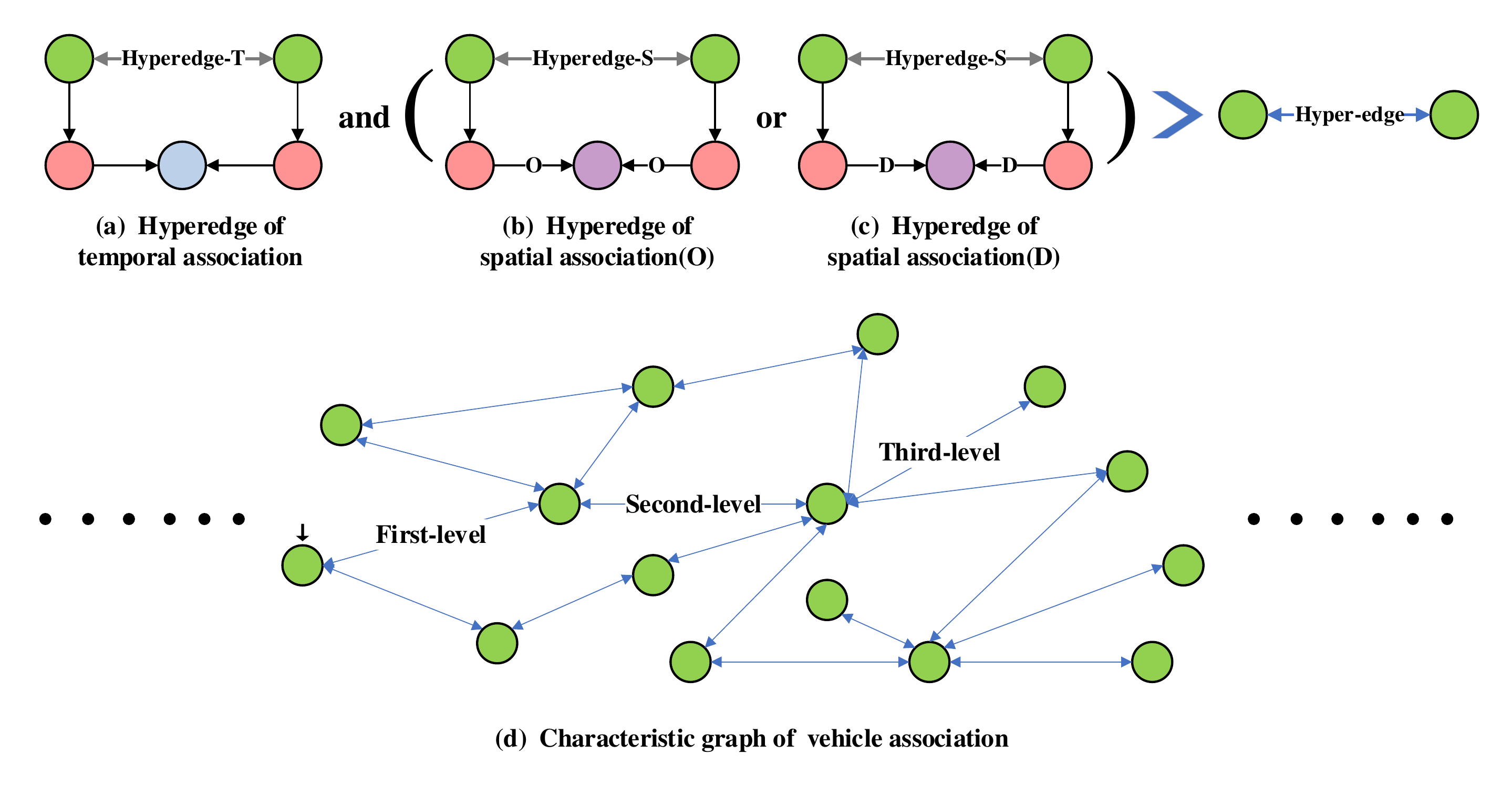}
\caption{Characteristic graph of vehicle association.}
\label{fig:cha_graph_ass}
\end{figure}

\begin{itemize}
    \item[$\bullet$] Temporal association: Trip at the same time span. Figure \ref{fig:cha_graph_ass}(a) shows the trip subgraph of two vehicle individuals with temporal association. We construct a super-edge that "temporal association" between them and use $v_m\stackrel{T}{\longleftrightarrow}v_n$ to indicate that vehicle entity $v_m$ and $v_n$ have temporal association.  
    \item[$\bullet$] Spatial association: Trip with the same origin or destination. Figure \ref{fig:cha_graph_ass}(b) shows the subgraphs of two vehicles' trips which have the same origin of a day, while the trips of two vehicles shown in Figure \ref{fig:cha_graph_ass}(c) have the same destination. For the vehicles $v_m$, $v_n$ whose trips has the above association are expressed by $v_m\stackrel{S-O}{\longleftrightarrow}v_n$ and $v_m\stackrel{S-D}{\longleftrightarrow}v_n$ respectively, witch belong to the super-edge of the spatial association. 
\end{itemize} 

On this basis, we constructed a  hyper-edge  ``$\stackrel{ST}{\longleftrightarrow}$" defined in Equation \ref{eq:ass} to represent the " hyper-association " between vehicles. Then the characteristic graph of vehicle depth association can be constructed as shown in \ref{fig:cha_graph_ass}(d). In this graph, there is only one type entity, and the all relations between $Vehicle$ entities is hyper-edge. Taking any vehicle entity as a base in the characteristic graph, we can get the vehicles it is associated to and the number of them by different levels. 
\begin{equation}\label{eq:ass}
    [(v_m\stackrel{S-O}{\longleftrightarrow}v_n \ or \ v_m\stackrel{S-D}{\longleftrightarrow}v_n) \ and \ v_m\stackrel{T}{\longleftrightarrow}v_n] \Rightarrow v_m\stackrel{ST}{\longleftrightarrow}v_n
\end{equation}

Assuming that the number of $Vehicle$ in the characteristic graph is $M$ , and considering the $N$-level association totally. Denote the number of vehicles associated with the $n$-level association of $m$-th vehicles $v_m$ can represented as $c_{mn}$. On this basis, the number of associated vehicles at all levels of each vehicles can be organized by a vector, that is, for the $m$-th vehicle, the number of associated vehicles at different levels can be expressed as $[c_{m1},c_{m2},\cdots,c_{mN}]$. On this basis, we can obtain the normalized vector $[r_{m1},r_{m2},\cdots,r_{mN}]$ of the vehicle by Equation \ref{eq:norm}. Define the mapping of vehicles to the normalized vector of associated vehicles number at each level as a function $A$, see Equation \ref{eq:A1}.  
Considering multiple vehicles in the characteristic graph, it can be modeled as a matrix, see Equation \ref{eq:A2}.

\begin{equation}\label{eq:norm}
    r_{mn} = \frac{c_{mn}}{\sum_{i=1}^N{c_{mi}}}
\end{equation}

\begin{equation}\label{eq:A1}
    A(v_m) = [r_{m1},r_{m2},\cdots,r_{mN}]^\mathrm{T}
\end{equation}

\begin{equation}\label{eq:A2}
\begin{aligned}
A(Graph) &= [A(v_1),A(v_2),\cdots,A(v_M)]^\mathrm{T} \\
        &=\begin{bmatrix}
r_{11} & r_{12}  & \cdots   & r_{1N}   \\
r_{21} & r_{22}  & \cdots   & r_{2N}  \\
\vdots & \vdots  & \ddots   & \vdots  \\
r_{M1} & r_{M2}  & \cdots\  & r_{MN}  \\
\end{bmatrix}
\end{aligned}
\end{equation}

For the matrix $A(Graph)$, the mean vector $v_{avg}=[r_1^{avg},r_2^{avg},\cdots,r_3^{avg}]$ can be calculated by Equation \ref{eq:mean}. 

\begin{equation}\label{eq:mean}
    r_j^{avg} = \frac{r_{ij}}{\sum_{i=1}^M{r_{ij}}}
\end{equation}

\subsubsection{Trip knowledge graph generation}

Through the previous work, we decompose the trip knowledge graph into a complete set of unit graphs. On this basis we extracted the information from these unit graphs and constructed the characteristic graphs. Theoretically, the characteristic graphs are constructed on the basis of a complete unit graphs set, that is, the information contained in the characteristic graphs is complete. So a trip knowledge graph should be able to be restored according to the characteristic graphs. Based on this idea, we study how to generate trip knowledge graph based on characteristic graphs we constructed. 

It should be noted that we cannot simply restore the information of the characteristic graphs totally, because this would make the generated trip knowledge graphs almost identical to the original ones, which would not solve the problem of data privacy. Therefore, how to combine each characteristic graph to design the logic of generation and how to design the randomness of generation are the difficulties of this study.

Before introducing the trip knowledge graph generation method, we first explain that each generation step is based on the date subgraph, i.e. the overall logic of generation is a rolling generation by date. Besides, for simplicity of presentation, we denote the four types of characteristic graphs in the order of introduction as $F_P$,$F_T$,$F_S$,$F_A$ respectively. And the trip unit graph is denoted as $u$.

The trip knowledge graph generation method proposed in this paper can be divided into the following three steps and each step will be described in detail next.
\begin{itemize}
    \item[(1)] Generation of trip unit graphs.
    \item[(2)] Determine the candidate trip unit graph combinations.
    \item[(3)] Select the optimal trip unit combination.
\end{itemize}

\paragraph{Generation of trip unit graphs}

\begin{figure}[h]
\setlength{\abovecaptionskip}{0.cm}
\centering
\includegraphics[width=12cm]{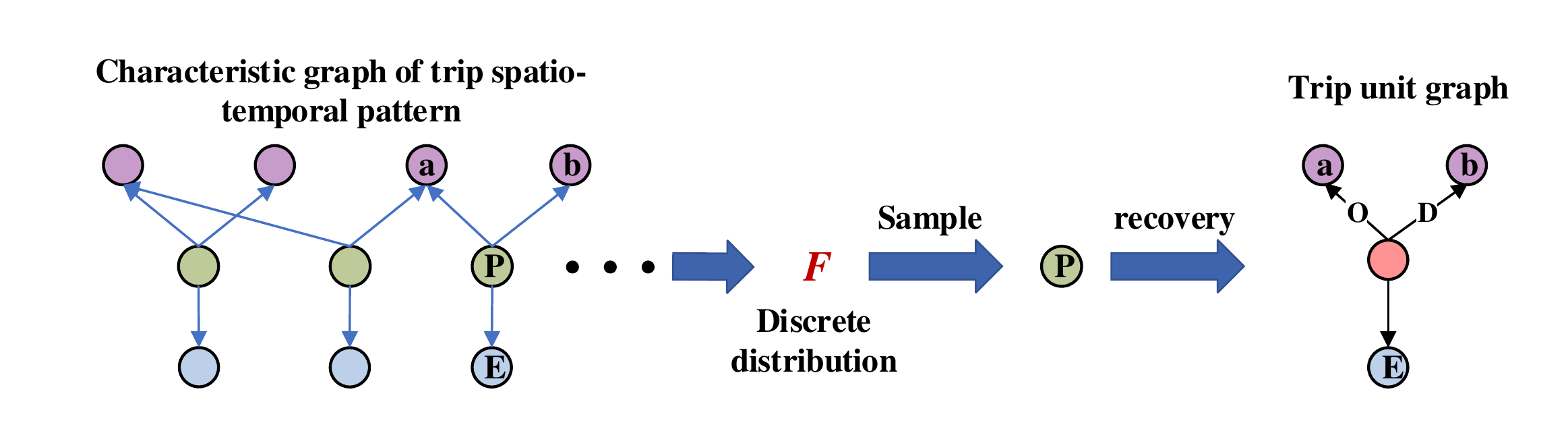}
\caption{Generation of trip unit graphs.}
\label{fig:generation1}
\end{figure}

The core component of the trip knowledge graph is the trip unit graph and we generate it firstly in the process of generating the trip knowledge graph. In constructing the characteristic graph, we constructed the characteristic graph of trip spatio-temporal pattern for the trip unit graphs and obtained a discrete distribution $F$ based on it. As shown in Figure \ref{fig:generation1}, the way we generate a single trip unit graph is as follows. Firstly,obtain a hyper entity of $Trip \ pattern$ type by sampling according to the discrete distribution $F$. Then extract the hyper entity and its association with $Zone$ and $TimeSpan$ entities from the characteristic graph. Finally,replace the hyper entity  by a $Trip$ entity and then a trip unit graph is gotten. If we need to generate $n$ trip unit graphs, then we just need to repeat the above process $n$ times.

Since the discrete distribution $F$ is obtained from the trip unit graphs of the original graph, this generation method of trip unit graphs can ensure the similarity between the generated trip knowledge graph and the original one in terms of trip spatio-temporal patterns. The process of this step can be described as algorithm \ref{alg:algorithm1}.

\begin{algorithm}[h]
 \caption{Generation of trip unit graphs}
 \label{alg:algorithm1}
 \KwIn{A date subgraph of original knowledge graph: $G^d$;Number of trip unit graphs generated: $n$ }
 \KwOut{The set of generated trip unit graph: $U$}
 Construct $F_P$ of $G^d$\;
 Get $F$ of $F_P$ by Equation \ref{eq:trip_pattern}\;
 $U \leftarrow \varnothing$\;
 \SetKwRepeat{Do}{do}{while}
 \Do{$\vert U \vert < n $ }{ 
 Sample $h$ from $F$ \;
 Generate $u$ according to $h$ \;
 $U \leftarrow U \cup \{u\}$\;
 }
\end{algorithm}

\paragraph{Determine the candidate trip unit graph combinations}

With the previous step, we got the trip unit graphs of all trips in a day. The following problem to be solved is how to combine and distribute them to vehicles. In other words, we need to determine the combination of trip unit graphs generated by the same vehicle and construct the relations of  $hastrip$ type between them. 

We take a single vehicle as a unit for the combination of trip unit graphs. For a specific vehicle, we  need to determine its candidate of trip unit graph combinations firstly. Assuming that we are generating the trips of the vehicle corresponding to entity $v$ on the day, the method for determining the combinations of candidate for that vehicle is given below. As shown in Figure \ref{fig:generation2}, we first obtain the vector $C(v)$ based on its characteristic graph of trip temporal combination in the original trip knowledge graph. On the other hand, we obtain its current position based on its characteristic graph of trip spatial continuity in which the long edge represents the origin and destination of trip that has been generated. The bottom of  figure \ref{fig:generation2} shows the generated trip unit graphs, which we grouped according to the entities of $TimeSpan$ type and sorted in the order of occurrence of the time. We can parse a combination of time spans from $C(v)$, and this information determines in which trip unit graphs we combine for that vehicle. For example, in this figure we can get three time spans, $A$, $B$, and $C$. Then we will choose one of the corresponding three types of trip unit graphs in turn to make the combination. The selection of the trip unit graph and the combination process need to be based on the principle of prioritizing the spatial continuity of the trip combination. With the above process, we can obtain the candidate combinations of trip unit graphs for the vehicle.

\begin{figure}[h]
\setlength{\abovecaptionskip}{0.cm}
\centering
\includegraphics[width=12cm]{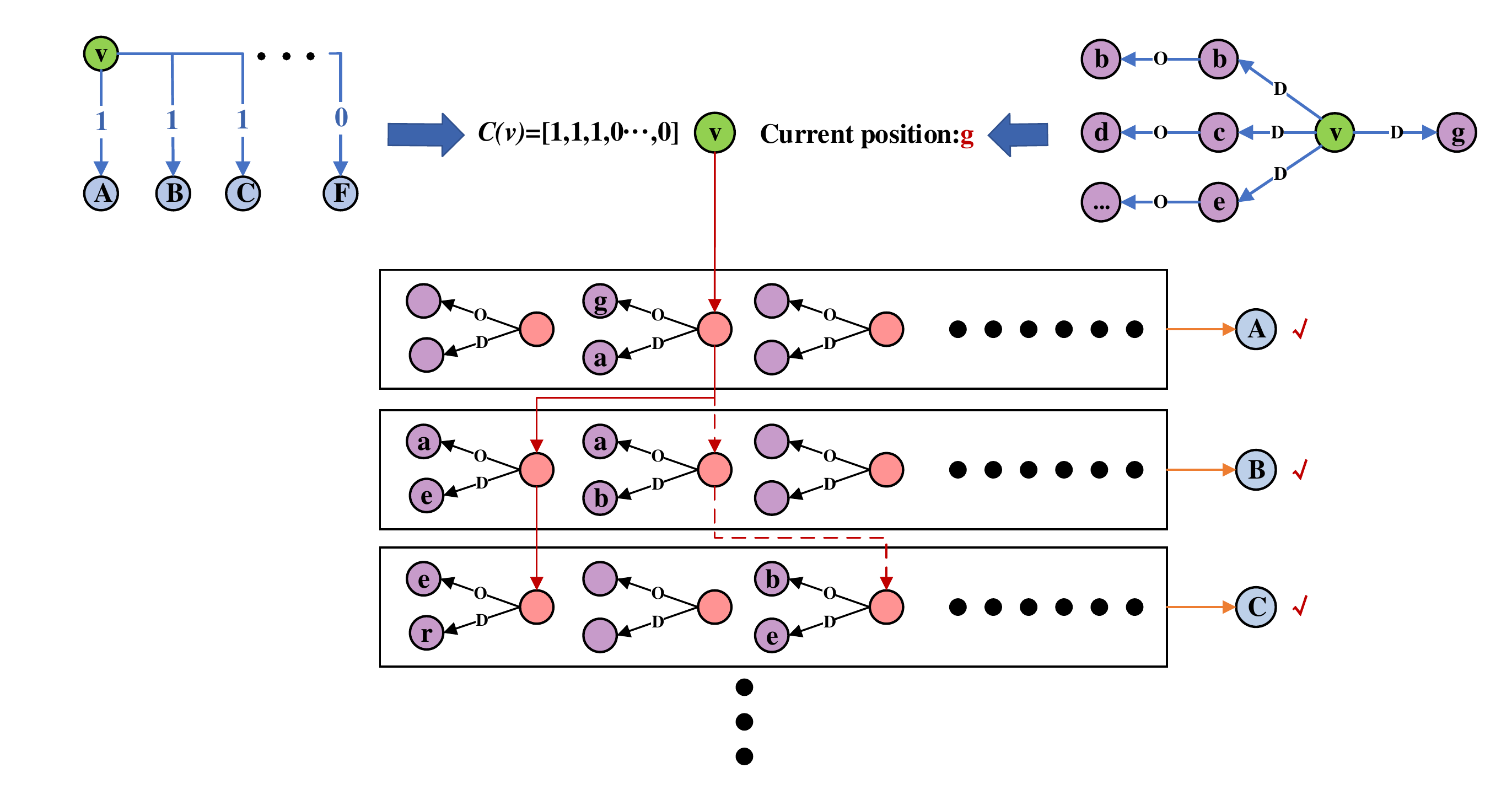}
\caption{Determine the candidate trip unit graph combinations.}
\label{fig:generation2}
\end{figure}

\paragraph{Select the optimal trip unit combination}

\begin{figure}[h]
\setlength{\abovecaptionskip}{0.cm}
\centering
\includegraphics[width=12cm]{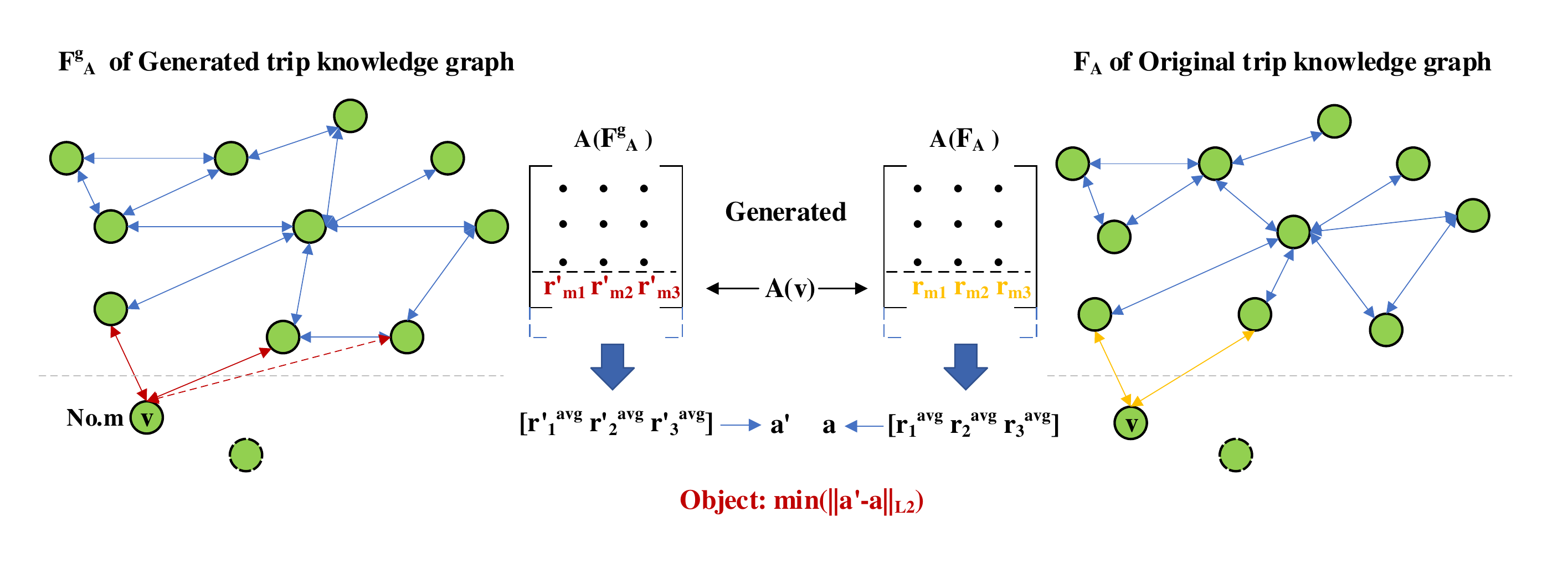}
\caption{Select the optimal trip unit combination.}
\label{fig:generation3}
\end{figure}

The combination of trip unit graphs obtained through the previous step is usually not unique, even the trip spatial continuity is fully guaranteed. Hence we need to choose an optimal combination among all the candidate trip unit graph combinations. In the section on extracting unit graphs it is introduced that the association between vehicles in the trip knowledge graph is generated through trips. Conversely, for vehicles, assignment to different trip combinations generates different associations between vehicles. As an example, in Figure \ref{fig:generation2}, the red solid and dashed lines are two candidate  combinations, respectively, and the corresponding associations between vehicles are illustrated in Figure \ref{fig:generation3} when vehicle $v$ selects them separately.

In summary, different combinations of trip unit graphs mainly affect the association between vehicles. We define the optimal trip unit combination as the trip combination that can make the vehicle association information of the generated knowledge graph most similar to the original trip knowledge graph. As shown in \ref{fig:generation3}, the left side is the characteristic graph of vehicle association among the generated vehicles of the generated knowledge graph, denoted as $F_{A}^{g}$, and the right side is the vehicle association characteristic graph of the same vehicles as the original trip knowledge graph, denoted as $F_{A}$. Assuming that we are selecting the $m$-th  vehicle's combination, then we take the vehicle into account in the characteristic graph. Hence the vehicle association matrix is $[A(v_1),A(v_2),\cdots A(v_m)]^\mathrm{T} $ and the expansion of two knowledge graphs is shown in Figure \ref{fig:generation3}. For the $m$-th vehicle, different combinations cause changes in the association matrix of the generated trip knowledge graph, while the association matrix of the original trip knowledge graph is stable. We can calculate the mean vectors of the two association matrices by Equation \ref{eq:mean}, denoted as $a'$ and $a$, respectively. Then the optimal combination can be described as a combination of trip unit graphs that can make the $F_{A}^{g}$ to satisfy the object that $min(||a^{'}-a||_{L2})$.

When the next vehicle combination is selected, we then add it to the association matrix. So this step is an online iterative optimization search process.

We denote that the set of candidate trip unit graph combinations of $v$ as $Y$. Then the algorithmic flow of this step is shown in Algorithm \ref{alg:algorithm2}. For the efficiency of algorithm execution it is also possible to set a very small deviation threshold to make the algorithm exit earlier. 

\begin{algorithm}[h]
 \caption{Select the optimal trip unit combination.}
 \label{alg:algorithm2}
 \KwIn{The original trip knowledge graph: $G$;The generated trip knowledge graph: $G^g$; $v$; $Y$}
 \KwOut{The optimal combination of trip unit graphs of $v$: $S$}
 $b_{min} \leftarrow \infty$\;
 $S \leftarrow \varnothing$\;
 Retrieve the set $V^g$ of $Vehicle$ from $G^g$ \;
 Extract $G_d^l$ from $G$ by constraining ``$Date<=d_i$ and $Vehicle$ $\in$ $V^g$" \;
 $M$ $\leftarrow$ $A(G_d^l)$\; 
 Calculate $a$ of $M$ by Equation \ref{eq:mean}\;
 \For{$y$ in $Y$}{
 Associate $v_i$ with $u$ of $y$ by $hastrip$ to get graph $G^a$\;
 Construct $F_A$ of $G^a$\;
 Calculate the mean vector $a^g$  of $A(G^g)$ \; 
 $b \leftarrow ||a-a^g||_{L2}$\;
 \If{$b<b_{min}$}{
 $b_{min} \leftarrow b$; $S \leftarrow y$\;}
  }
\end{algorithm}

\section{Experiments} \label{sec:exper}

\subsection{Data Description}

In this study, 35 days of trip data of Xuancheng city ranging from $01/08/2019$ to $04/09/2019$ were used which is collected by automatic vehicle identification (AVI) systems. Through the data prepossessing such as trip classification and exclusion of operating vehicles, the data set includes a total of 8,300,091 trips from 822,569 vehicles. The fields contained in each of these records and their meanings are shown in Table \ref{tab:field of trip records}. 

\begin{table}[h]
\centering
\caption{Fields of vehicle individual trip records}
\label{tab:field of trip records}
\begin{tabular}{l|l}
\hline
\textbf{Field}      & \multicolumn{1}{c}{\textbf{Description}}      \\ 
\hline

Vehicle & Unique identification of individual. \\
Ftime    & The beginning time of, e.g.$09:30:00$.   \\
Date    & The date of trip, e.g. $2019-08-01$.   \\
Fzone & The origin zone of trip.  \\
Tzone    & The destination zone of trip.           \\ \hline

\end{tabular}
\end{table}
Based on the trip data in Table \ref{tab:field of trip records}, the individual-level trip knowledge graphs can be constructed according to the structure introduced in Figure \ref{fig:structure_KG}, in which The  ``$Week$" and ``$TimeSpan$" entities  are the mapping of the ``$Date$" field and the ``$Ftime$" field in Table \ref{tab:field of trip records} respectively.

The amount of both entities and relations of the individual-level knowledge graph reaches over 10 million.

\subsection{Evaluation of individual mobility characteristic mining}

Based on the constructed individual-level trip knowledge graph, we are able to divide all vehicles into five groups, such as commuters. The percentage of vehicle number and their trip number of different labels is shown in Table \ref{tab:eva trip characteristic}. 

\begin{table}[htb]
\centering
\caption{Individual trip characteristics mining results}
\label{tab:eva trip characteristic}
\begin{tabular}{lcc}
\hline
\multicolumn{1}{c}{Trip type} & Percentage of  vehicle number & Percentage of  trip number \\ \hline
Passing vehicle               & 81.07\%                       & 12.19\%                    \\
Commuter                      & 1.51\%                        & 11.38\%                    \\
Vehicle of random             & 15.44\%                       & 58.83\%                    \\
Vehicle of stable             & 1.52\%                        & 5.00\%                     \\
Vehicle of high   frequency   & 0.46\%                        & 12.60\%                    \\ \hline
\end{tabular}
\end{table}

\begin{figure}[h]
\centering

\subfigure[Workday]{\includegraphics[width=0.49\textwidth]{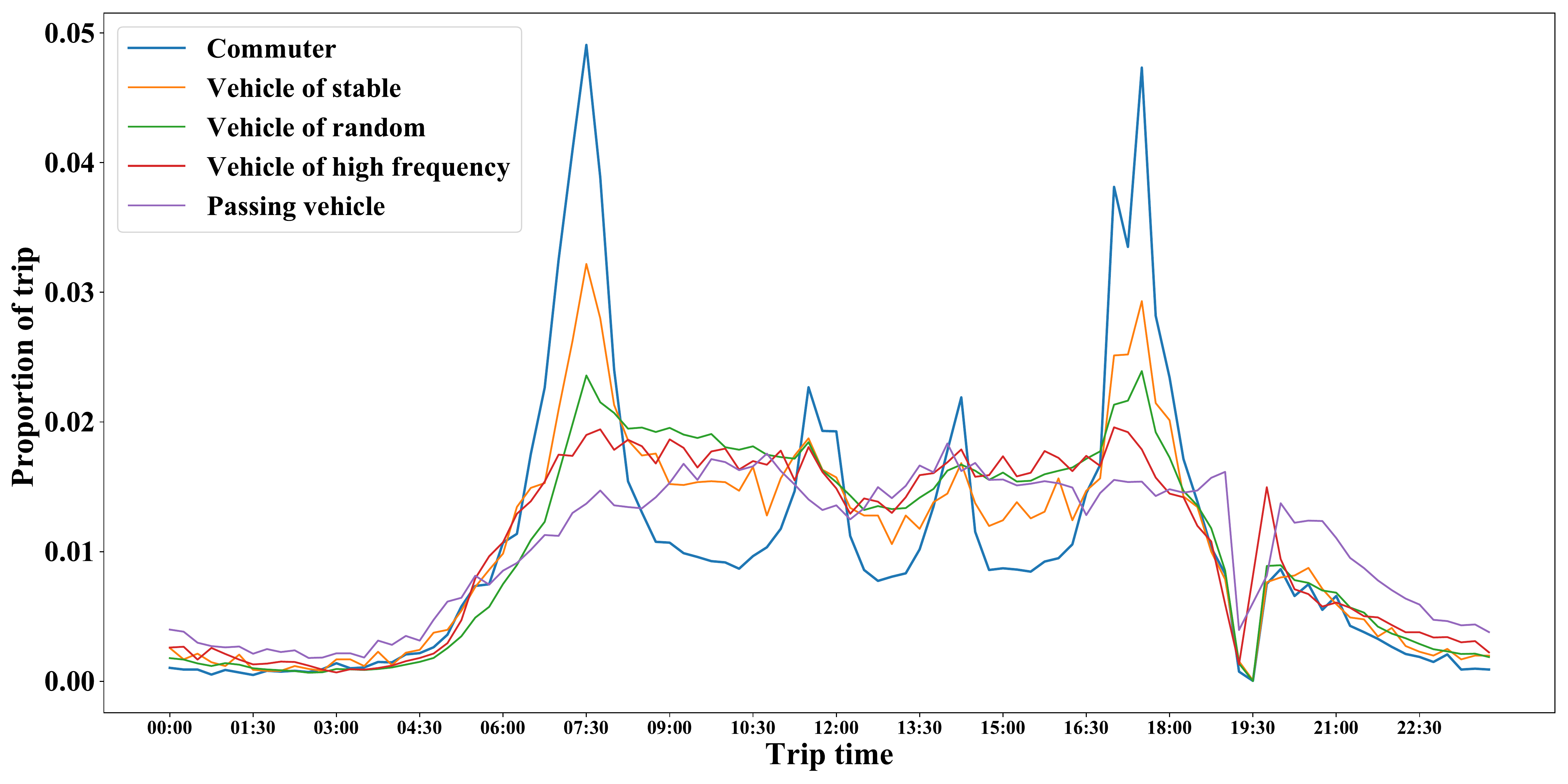}}
\subfigure[Holiday]{\includegraphics[width=0.49\textwidth]{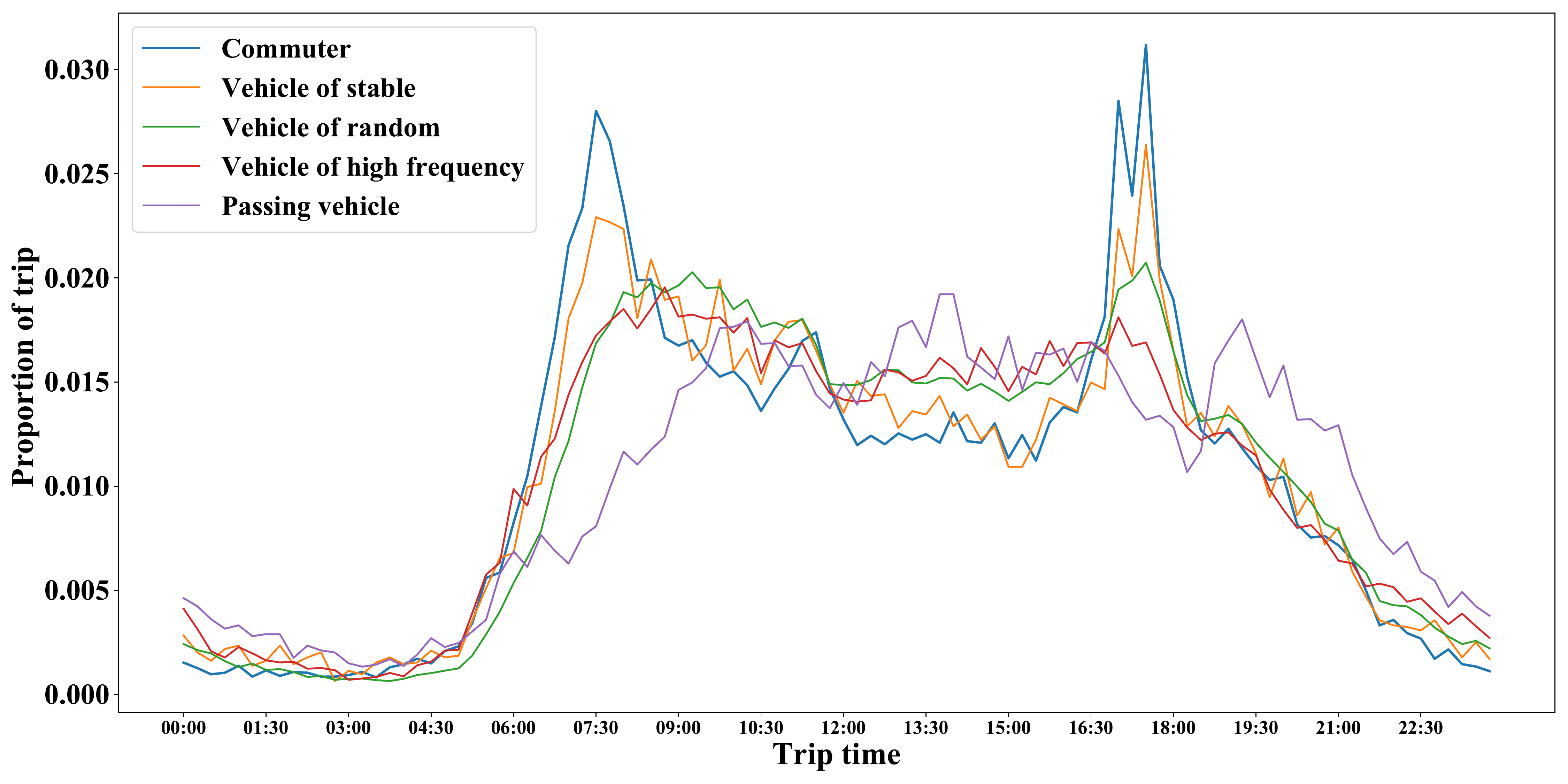}}
\caption{Trip number distribution with time of vehicles  with different label }
\label{fig:time_dis}
\end{figure}

In order to demonstrate the effect of individual mobility characteristics mining from temporal perspective, we draw the distribution of proportion of trip number of  vehicles that have different label with time at a granularity of 15 minutes, see Figure \ref{fig:time_dis}. It can be seen that the commuters delineated by mining has obvious commuting characteristics on workdays, i.e., there are obvious morning peak, evening peak and lunchtime peak phenomena. Compared to commuters, trips of vehicle of random and other labels are not exhibit significant temporal patterns.

From the spatial perspective, we selected commuters' high trip-prone zones and analyzed them with the city's POI information, and found that these zones cover POIs of residential, school office buildings, hospitals and other commuting scenarios.

\subsection{Evaluation of the trip generation algorithm}
We have generated five label subgraphs using Algorithm \ref{alg:algorithm1}, which compose a complete individual-level trip knowledge graph. Hence we obtained the generated trip data by retrieving from generated trip knowledge graph. The generation of trip knowledge graph is based on the label subgraph, so the generated data can be matched to a specific group of vehicle. For the trip data of each group of vehicle, we evaluate it from three aspects: trip spatio-temporal pattern, vehicle association and trip spatial continuity.

\subsubsection{Evaluation of trip spatio-temporal pattern}

For this item, we mainly compare the similarity of trips in temporal and spatial distribution between the generated trip data and historical trip data. The KL divergence is introduced to evaluate  distribution and the KL divergence of different groups of vehicle between historical trip data and generated trip data is shown in Table \ref{tab:KL_ts}. For the spatio distribution, the KL divergence is calculated by overall O-D combination regardless of time factor.

\begin{table}[htb]
\centering
\caption{KL divergence of trip temporal and spatial distribution (generated v.s. historical).}
\label{tab:KL_ts}
\begin{tabular}{l|c|c|c|c}
\hline
          &Commuter & Vehicle of stable & Vehicle of random & Vehicle of high frequency \\ \hline
 Temporal &0.00017  & 0.00023           & 0.00054           & 0.00030  \\          
 Spatial &0.055    & 0.067             & 0.035             & 0.077       
 \\\hline
\end{tabular}
\end{table}

\begin{figure}[h]
\centering
\subfigure[Commuter]{\includegraphics[width=0.48\textwidth]{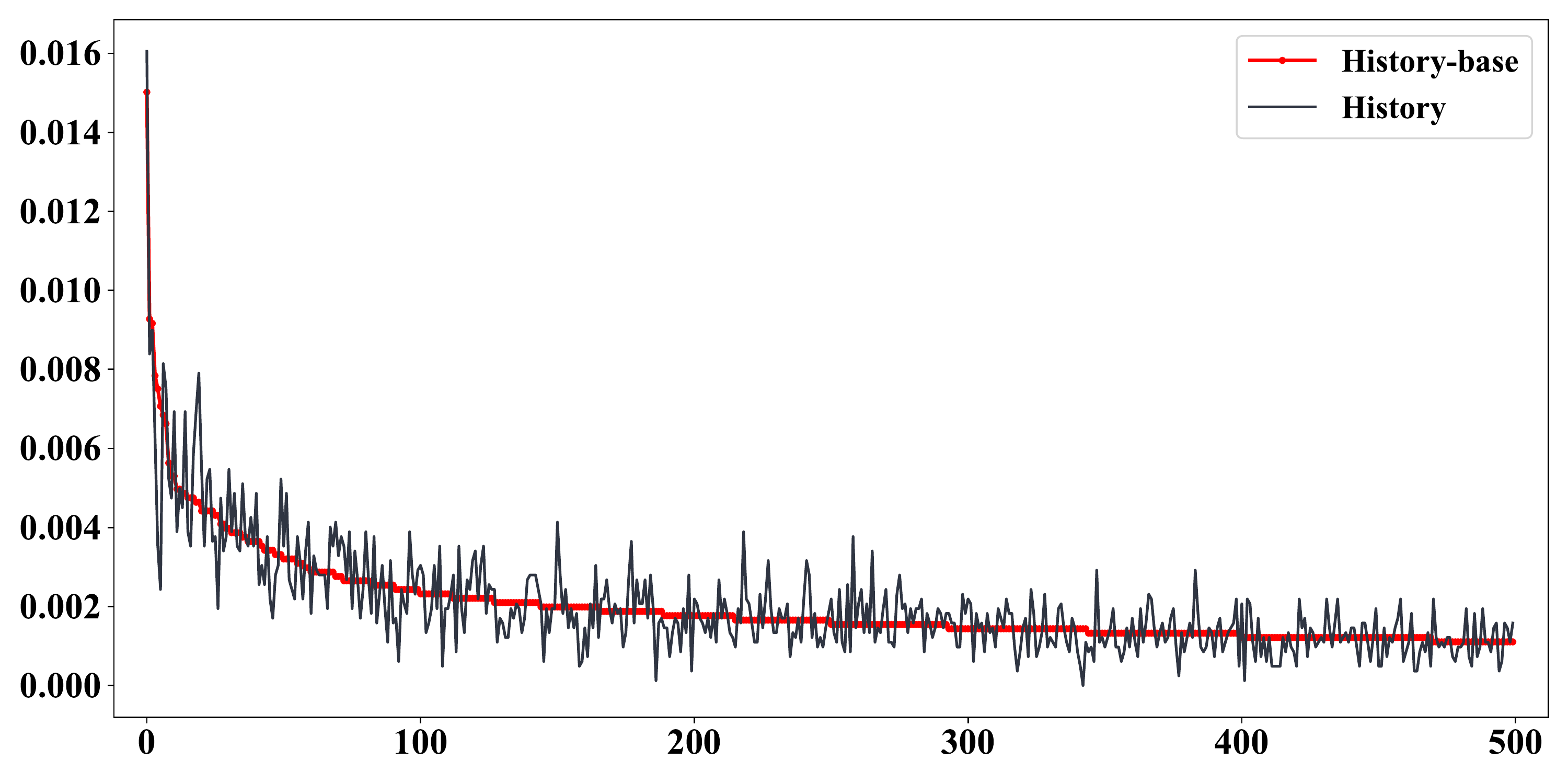}}
\subfigure[Commuter]{\includegraphics[width=0.48\textwidth]{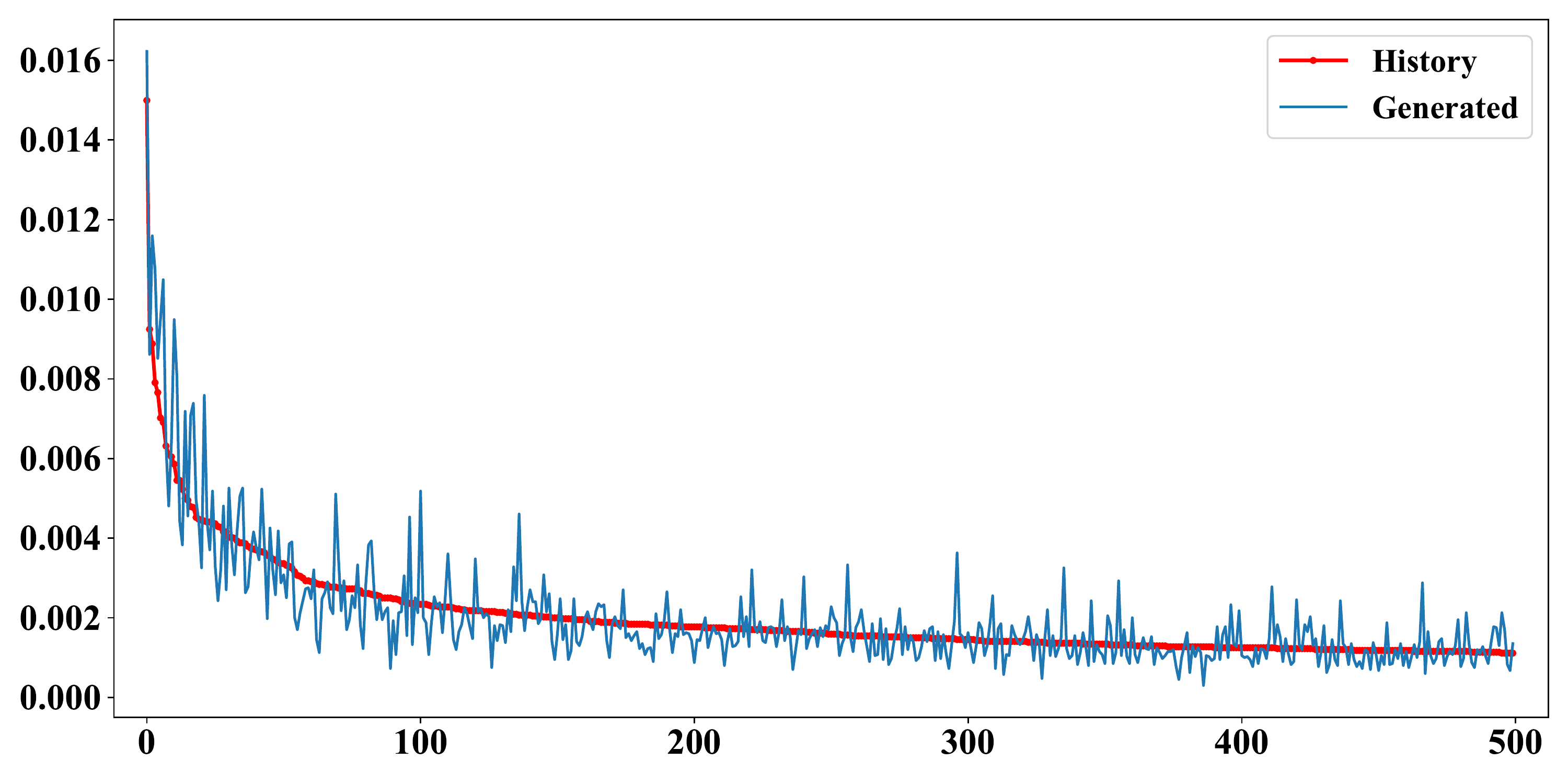}}
\subfigure[Vehicle of random]{\includegraphics[width=0.48\textwidth]{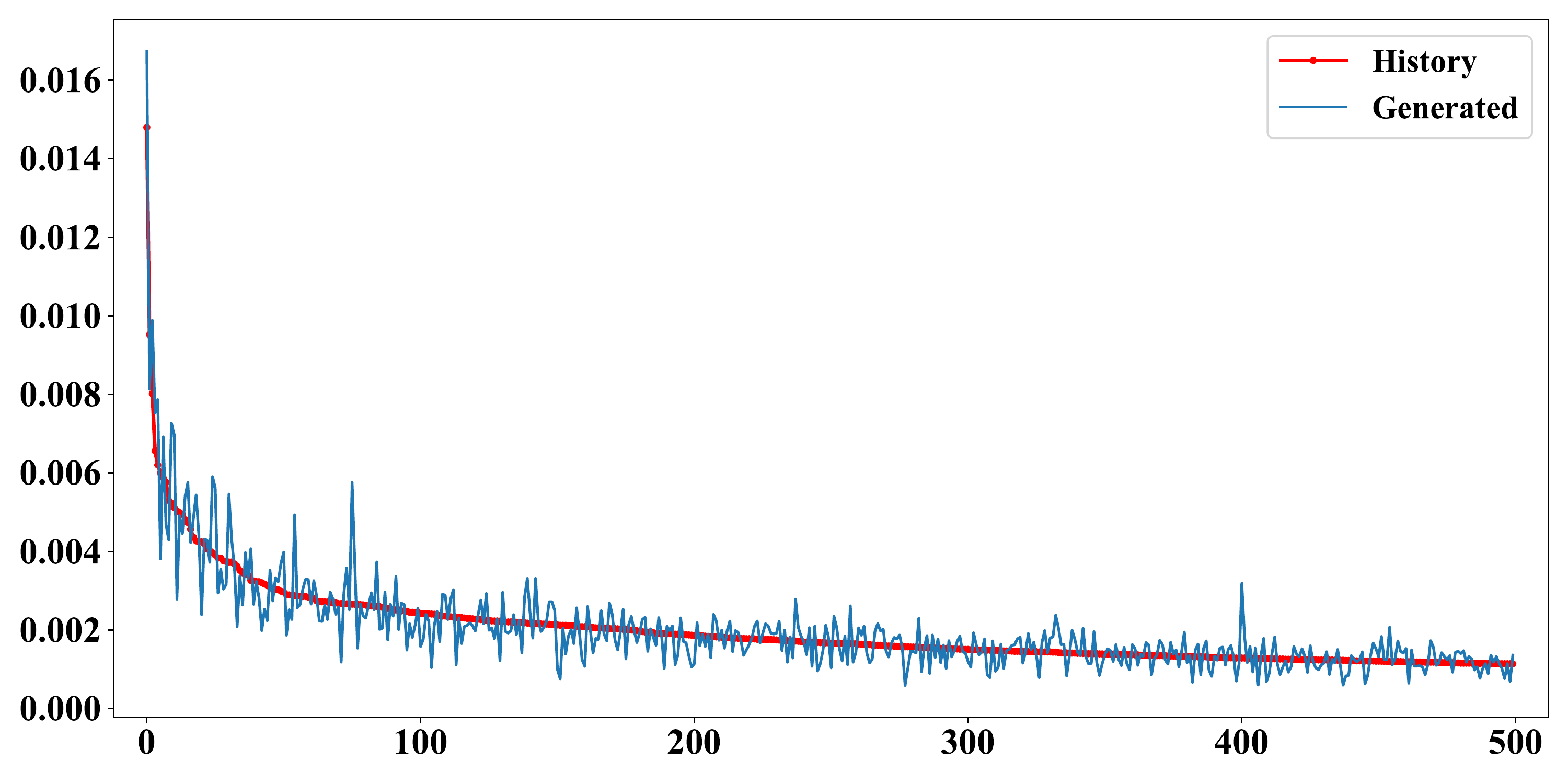}}
\subfigure[Vehicle of high frequency]{\includegraphics[width=0.48\textwidth]{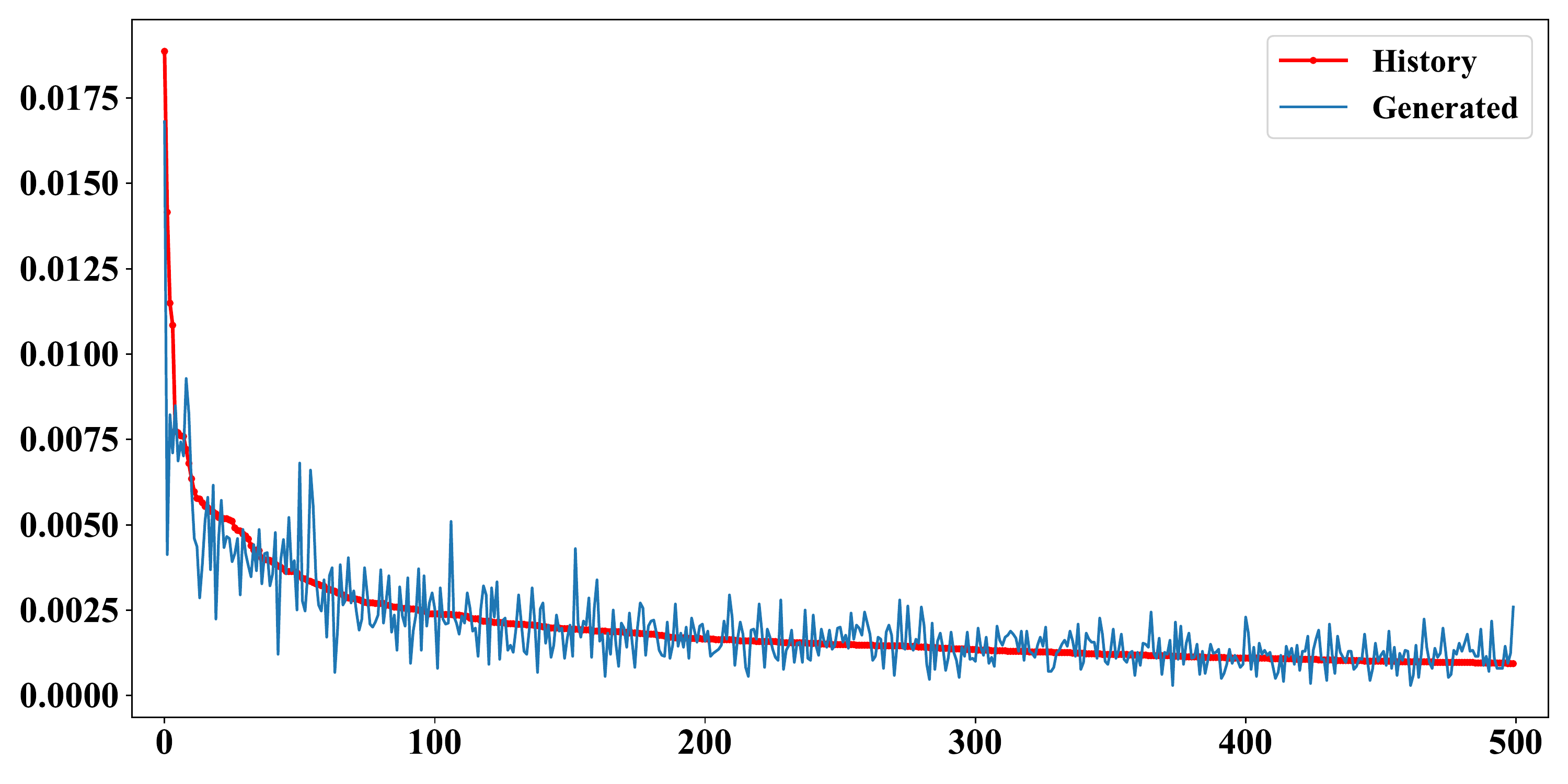}}
\caption{The distribution of top-500 OD-flow.}
 \label{fig:OD-flow distribution}
\end{figure}

In order to show the correlation between the KL divergence and the distribution similarity visually, we plotted the flow distribution curves of  main ODs of partial vehicle types,see Figure \ref{fig:OD-flow distribution}. In this figure, The horizontal axis corresponds to the different OD combinations, which we use their descending ranking under the benchmark(red curve) to represent them. The vertical axis is the proportion of trips of the corresponding OD combinations. Figure \ref{fig:OD-flow distribution} (a) shows the fluctuations between two weeks of historical data for commuters, using one of those weeks as a benchmark.

As shown in Table \ref{tab:KL_ts}, we can see the temporal  and spatial KL divergence of different groups of vehicle between generated data and historical data are on the order of $10^{-4}$ and $10^{-2}$ respectively, which are same as the normal fluctuation between historical data. Besides,  Figure \ref{fig:OD-flow distribution} shows the distribution of flow of the top 500 OD combination. It visually shows the similarity of the differences in OD flow distribution between historical data and between historical data and generated data.

\subsubsection{Evaluation of vehicle association}

In the actual generation process, we considered a total of three levels of vehicle association. Through the generation we found that, if we construct the characteristic graph of vehicle association with same number of vehicles, the proportion in the number of vehicles associated with each level is stable of each group, while there is a large difference between different groups. Table \ref{tab:ass_tab} shows the proportion at each level for different groups of vehicles, and the deviation of generated and historical data is shown in last column. The calculation of deviation is shown in Algorithm \ref{alg:algorithm2}.

\begin{table}[htb]
\centering
\caption{The proportion of the number of vehicle associated with three level.}
\label{tab:ass_tab}
\begin{tabular}{lcccc}
\hline
                         & First level & Second level & Third level & Bias\\ \hline
Commuter                 & 0.05        & 0.32         & 0.63        &0.0724\\
passing car              & 0.28        & 0.33         & 0.39        &0.0652\\
vehicle of high frequncy & 0.14        & 0.43         & 0.43        &0.0634\\
vehicle of   stable      & 0.06        & 0.28         & 0.66        &0.0732\\
vehicle of   random      & 0.06        & 0.34         & 0.60        &0.0806\\ \hline
\end{tabular}
\end{table}

\subsubsection{Evaluation of trip spatial continuity }

We calculated that the spatial continuity of historical trips for different vehicle groups is between 60\% and 70\%. The trip generated spatial continuity of each groups of vehicle is shown in Table \ref{tab:Continuity}. It can be seen that the spatial continuity of the generated trip data is better than the historical trip data, thus further enhance the usability of the generated trip data for different academic and engineering application.

\begin{table}[htb]
\centering
\caption{The spatial continuity of trip data generated.}
\label{tab:Continuity}
\begin{tabular}{c|c|c|c|c}
\hline
 Commuter & Vehicle of stable & Vehicle of random & Vehicle of high frequency& Passing vehicle \\ \hline
 93.1\%  & 92.4\%          & 91.4\%          & 85.6\% &     95.4\%            \\ \hline
\end{tabular}
\end{table}

\subsection{Discussions on generated data}

Utility and privacy are competing factors, and the trip data generated using the method proposed in this paper will have some compromise in utility, although the privacy of individuals and is protected. This section focuses on the compromise in utility of the generated data.

First, we can see  that the generated data are able to maintain a high degree of similarity with the historical data in the spatio-temporal patterns of groups macroscopically through the experimental part. Therefore, for the macro aspect of the task, there is no compromise in the trip data generated in this paper. At the individual level, the trip temporal combination pattern characteristic guarantees the trip frequency of individual and trip period combinations. Hence the generated data is equally uncompromising for analyzing the distribution of trip frequency and trip periods of individuals in the city. As for the spatial information of individual trip, since the method proposed in this study protects privacy by disorganizing the destination selection of individuals, for each individual, its trip activity chains in space is distorted. Therefore, if such a study like analysing the similarity of traffic zones by the preference of individual choice of destination is to be conducted, the results of using generated data will differ significantly from the historical data.

\section{Conclusion}

In this study, we focus on solving the problem of organizing, analyzing and especially in generating urban individual-level trip data by introducing knowledge graphs. First, we designed and constructed an individual-level trip knowledge graph. It expresses individual-level trip data in the way of entity and relation, which greatly improves the efficiency of obtaining individual-level trip data and it well expresses the association information. Secondly, we mined the trip characteristics for each vehicle based on the individual-level trip knowledge graph. On this basis, all of urban vehicles are subdivided into  five groups. Finally, we proposed a trip generation method based on trip knowledge graph. This method can generate individual-level trip knowledge graphs by generating trip knowledge graph. The experiment shows that the final generated trips are similar to the historical trip data in trip patterns and vehicle associations, and have high spatial continuity.

\section{Acknowledgment}
This research is supported by the project of National Natural Science Foundation of China (No. U1811463).

\newpage


\bibliography{mybibfile}

\end{document}